# Non-Banking Sector development effect on Economic Growth. A Nighttime light data approach.

Leonard Mushunje[*1] and Maxwell Mashasha[2]

leonsmushunje@gmail.com

[1]Columbia University

[2]Midlands State University

**Abstract**

This paper uses nighttime light(NTL) data to measure the nexus of the non-banking sector, particularly insurance, and economic growth in South Africa. We hypothesize that insurance sector growth positively propels economic growth due to its economic growth-supportive traits like investment protection and optimal risk mitigation. We also claim that Nighttime light data is a good economic measure than Gross domestic product (GDP). We used weighted regressions to measure the relationships between nighttime light data, GDP, and insurance sector development. We used time series South African GDP data collected from the World Bank for the period running from 2000 to 2018, and the nighttime lights data from the National Geophysical Data Centre (NGDC) in partnership with the National Oceanic and Atmospheric Administration (NOAA). From the models fitted and the reported BIC, AIC, and likelihood ratios, the insurance sector proved to have more predictive power on economic development in South Africa, and radiance light explained economic growth better than GDP and GDP/Capita. We concluded that nighttime data is a good proxy for economic growth than GDP/Capita in emerging economies like South

Africa, where secondary data needs to be more robust and sometimes inflated. The findings will guide researchers and policymakers on what drives economic development and what policies to put in place. It would be interesting to extend the current study to other sectors such as micro-finances, mutual and hedge funds.

**Keywords:** Nighttime lights data, GDP/Capita, Radiance light, Saturated light, Economic growth, Insurance sector, Insurance Premium density.

## 1. Introduction

Before drawing the line between our contribution and the existing literature, we give our main contributions to this study as follows:

I. Measuring economic growth contribution from the insurance industry using satellite imagery and nightlight data is new, to the best of our knowledge.

II. The approach is convenient because all the daytime satellite images and nightlight intensity data are evenly accessible from the open-source online interfaces linked to South Africa.

III. The study approached the problem more robustly. We matched all the grid images of different economic sectors (insurance, micro finances) to South African country codes to generate corresponding features. Thus, ensuring less variation and noise within our data, analytics, and results.

The role of the non-banking sector in economic growth and development is considerably growing and is paid attention to across the globe. Nevertheless, scant research has been conducted to unravel the relationship between the insurance sector and real-time economic growth. It is vital to indicate that the insurance sector plays an imperative role in any economy by providing indemnity or investment protection. Moreover, its ability to pool funds in the form of premiums enables it to be a crucial institutional investor in emerging economies like South Africa. From another position, measuring economic activities and output is primarily done using traditional means of averaging Gross domestic product (GDP). However, this carries some bias, leading to incorrect and less accurate results and policies. Therefore, this study aims to measure the insurance-growth nexus in South Africa using nighttime light data (NTL) rather than Gross domestic product. Our conjecture is that; insurance sector businesses positively drive economic growth due to their economic growth-supportive Key Performance Indicators (KPIs) like investment protection and labor (life) insurance, among others. We also claim that Nighttime light data is an excellent economic growth measure than Gross domestic product.

Of further interest, the studies on the subject have been mainly of a cross-sectional or panel nature, for instance, Azman-Saini and Smith (2011) and Moshirian, et al. (2010). Most research done on the subject mainly employed a panel data analysis approach. The most significant drawback of this method is losing specific effects in analysis, which, if combined with inaccurate GDP data, needs to be more accurate. Therefore, it is essential also to interrogate the relationship between insurance sector development and economic growth using nighttime light data on a time series basis. In this study, we are using South Africa among all other emerging economies because of its stage of development and financial stability. Establishing the nature of the relationship between the insurance sector and economic growth in South Africa being our first intention, the performance comparative hypothesis testing for GDP and NTL data makes up the novelty of this paper. There is substantial literature explaining the role of the insurance sector in economic growth in South Africa.

Sibindi and Godi (2014) discussed the relationship between insurance growth and economic growth in South Africa. They used insurance density as the proxy for insurance market development and real per capita growth of domestic products for economic growth. They tested for cointegration amongst the variables by applying the Johansen procedure and then tested for Granger causality based on the vector error correction model (VECM). They found at least one co-integrating relationship and indicated that the direction of causality runs from the economy to the long-term insurance and from the economy to the entire insurance sector. Ward and Zurbruegg (2000) examined the relationship between economic growth and growth in the insurance industry for nine OECD countries. Using annual data, they conducted a bivariate cointegration analysis and tested for causality using the real GDP and the total real premiums in each country from 1961 to 1996. They found that in some countries, the insurance industry Granger causes economic growth; in others, economic growth Granger causes the development of the insurance sector. This agrees with the works of Haiss and Sümegi (2008), who investigated the impact of insurance investment and premiums on GDP growth in Europe using a cross-country panel data analysis for 29 European countries from 1992 to 2009.

Chi-Wei, Hsu-Ling, and Guochen (2013) also apply the bootstrap Granger causality test to examine the relationship between insurance development and economic growth in seven Middle Eastern countries. They as well used insurance density as the indicator for insurance development. Their results found evidence for bi-directional causality between the life insurance sector and economic growth in higher-income countries such as United Arab Emirates, Kuwait, and Israel. Their results agree with Chang, Lee, and Chang (2013), who studied the relationship between insurance and economic growth by conducting a bootstrap panel Granger causality test using data from 10 OECD countries throughout 1979-2006. Other related works include Ćurak et al. (2009), Moshirian, et al. (2010), and Ching et al. (2010), and most recently, Wang et al (2022); Pata, U. K., & Samour, A. (2022), and Ozturk, I., & Ullah, S. (2022). Also, Arena (2008) examined the causal relationship between insurance market activity and economic growth in developed and developing countries. The author employed insurance penetration (insurance premiums as a proxy for market development), and he found a positive relationship between the insurance activities and economic growth-insurance activities drive economic development.

Furthermore, Madukwe and Anyanwaokoro (2014) investigated the causality of the relationship between the life insurance business and economic growth in Nigeria from 2000-2011. Pearson's Product Movement Correlation Coefficient was used to test the hypothesis to determine the extent of the causality of the relationship between a life insurance business and economic growth. Their study revealed a significant causal relationship between the life insurance business and the economic growth of Nigeria. Additionally, Kugler and Ofoghi (2005) used the components of net written insurance premium to evaluate a long-run relationship between development in insurance market size and economic growth by using Johansen's $\lambda Trace$ and $\lambda max$ cointegration tests in addition to Granger causality tests. Using insurance penetration as a measure of insurance to economic growth, Olayungbo and Akinlo (2016) found a positive relationship for Egypt, while short-run adverse and long-run positive effects are found for Kenya, Mauritius, and South Africa. On the contrary, adverse effects are found in Algeria, Nigeria, Tunisia, and Zimbabwe. Mushunje and Mashasha (2020) looked at the role of the non-banking sector in the wall streets of the South African economy. Their concluded that the non-banking sector is indeed driving economic growth in South Africa, but they employed GDP as a measure of economic growth. Moreover, Mushunje and Mashasha (2020) looked at the growth patterns of the insurance sector in South Africa using a Gompertz growth model. They indicated positive growth in the sector since 2010 in South Africa.

All these works were based on investigating the role and causal links between insurance and economic growth, and their results proved to move in the same direction. They all found a positive link between economic growth and insurance businesses. However, despite all the interesting results from the literature, we found no direct application of nighttime light data as a proxy measure of economic growth when assessing the role of insurance in economic development. Only GDP was used to measure economic growth, which forms our paper's novelty. Application of nighttime light data appeared in literature but in other versions. For example, Zhang and Guo et al. (2019) compared the LJ1-01 nighttime light data with Visible Infrared Imaging Radiometer Suite (VIIRS) data in modeling socio-economic parameters. They used ten parameters from the four aspects of the gross regional product (annual average population, electricity consumption, and area of land in use) were selected to build linear regression models from the selected regions of China. The results showed that nighttime light data offered a better potential for modeling socio-economic parameters than the comparable VIIRS data; the former can be an effective tool for establishing models for socio-economic parameters. In addition, Mellander and Lobo et al. (2016) used a combination of correlation analysis and geographically weighted regressions to examine if light can function as a proxy for economic activities at a finer level. They used a fine-grained geo-coded residential and industrial full sample micro-data set for Sweden and matched it with radiance and saturated light emissions. Their results showed a correlation between NTL and economic activity and are strong enough to make it a relatively good proxy for population and establishment density. However, the correlation is weaker concerning wages. All these related works proved that nighttime light data (NTL) is a proxy measure of economic activities in an economy. However, they loosely looked at the association of NTL data as a proxy of economic growth and insurance businesses in South Africa.

## 1.1. Nighttime light(NTL) data NTL

It is a proxy for several variables, including urbanization, density, and economic growth. It is a better accurate measure of economic activity as it focuses on active and productive economic positions. Nighttime lights can be better economic measures in developing economies such as South Africa as such countries are the best models with almost unstable and less structured economies. Measuring and tracking how much productivity increases in a place is often more challenging. It becomes even more challenging in countries with poor national, let alone sub-national, statistics. In such countries, it is difficult to tell where people live and how fast the population grows. It is even harder to answer relevant policy questions regarding urban planning and transportation needs. Night Lights data can help us find answers to these questions. All these difficulties in the spotlight are pervasive in developing, and emerging economies such as South Africa, and nighttime light data is considered a better proxy measure for such.

Night Lights data was introduced in economics by researchers such as Henderson et al. (2012). Their work showed that nightlight data could supplement economic activity measures in countries with poor national statistics (emerging economies). They also give insights on developments in GDP and population on the sub-national level. Additionally, we see some related work on applying nighttime light data to economies in Pinkovskiy and Salai-Martin (2016) and Pinkovskiy (2016). We extracted a photograph illustrating the nighttime lights in Africa, as infigure1. The figure is clearly showing the light concentrations on a geographical basis. The light concentration is used to trace and measure economic activities. People use light to produce, construct and live in the modern world. The patterns of light, and the movement in these patterns over time, can guide us as to where people choose to live and produce and how their choices change. See below.

**Light concentration and patterns on a geographical basis in Africa.**

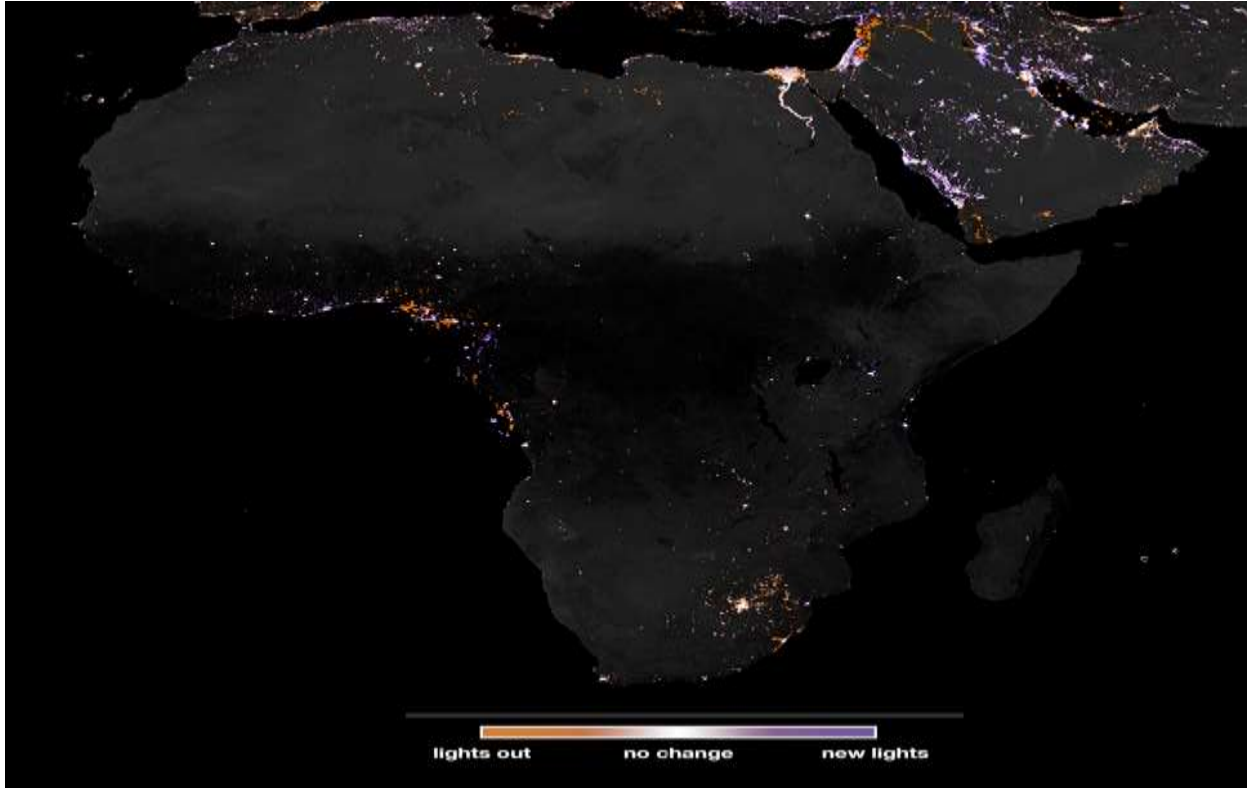

Figure 1: Source: NOAA's National Geophysical Data Center with author's additional remarks.

We have three usable nighttime light data: Stable lights ("Average Visible, Stable Lights, & Cloud Free Coverage"), unfiltered lights, and radiance-calibrated lights. In most cases, stable lights are used as they give better estimates and show the luminosity of cities and towns for more extended periods of interest. We shall, in this study, use stable nighttime lights data for South Africa as a proxy measure of Economic growth in the exploration of the role of the insurance sector on economic development. Below are additional examples of Satellite images of Nighttime Lights in Africa.

**Satellite images from Africa (1992 and 2013)**

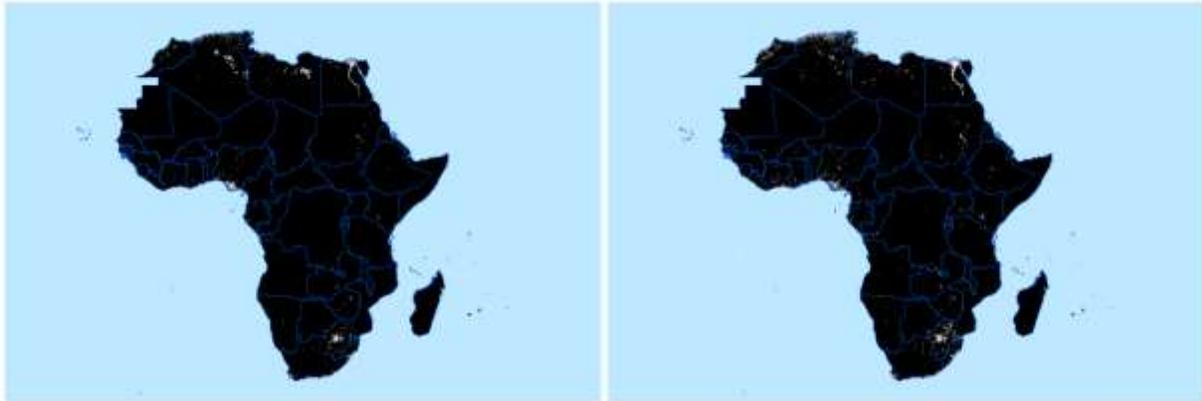

Figure 2: Source: IMF (2019) with authors' additional remarks.

We present examples of satellite images for the period 1992 and 2020, respectively, in Africa. Figure 2 gives a clear clue about the satellite and nighttime light data about South Africa- our country case. Different satellite systems are used to record the datasets of nighttime lights for extended analysis and comparative analysis with the real GDP.

### 1.2. Per Capita Gross domestic product (GDP/Capita)

It is a measure of total output produced within the borders of an economy per head, OECD (2023). It is considered a good measure of the economic growth of a nation. It is typically measured using three distinct methods: output, expenditure, and income. All of these give reasonable estimates if used correctly and appropriately, and the integrated use of all these is our recommendation. GDP is a good measure of economic growth only in countries where statistics are well-enriched and well-defined in a structured way. This is a common phenomenon in developed or well emerged economies. As such, we conjectured that nighttime light data is instead of GDP a good economic measure in developing and emerging economies like South Africa. GDP per capita is an important indicator of economic performance and a valuable unit for cross-country comparisons of average living standards and economic well-being. It is usually a non-static variable, meaning it varies with time. This may come from new resource discoveries or exhaustion, technological advancement, or a change in any factor that affects production. For South Africa, Figure 3 depicted below shows the GDP growth patterns.

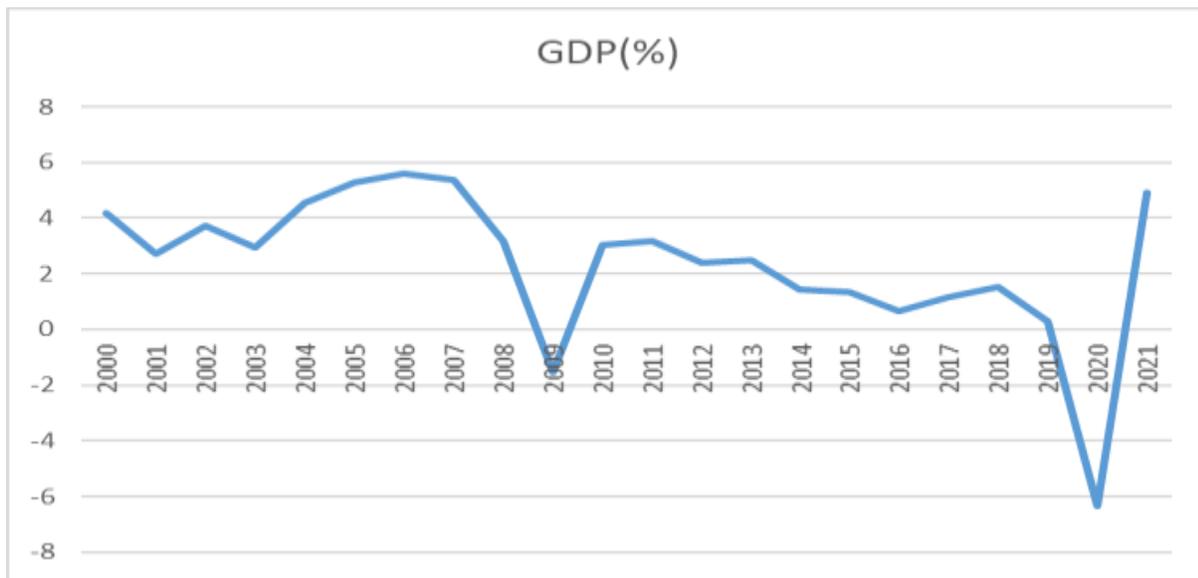

Figure 3:we present the GDP values (%) using World Bank data (2022). Ideally, the GDP fluctuates over the entire period considered. Interestingly, there was a sharp increase between 2019 and 2021 due to Covid19.

Concerning GDP per capita, there are other measures of economic growth, such as Gross national income (GNI), Real GDP, and Gross national product (GNP). These are all suitable measures of economic growth in countries with excellent and well-organized statistical figures, unlike in developing countries like South Africa, where we have the opposite case. Therefore, alternative methods, such as nighttime light data, are ideal in such countries.

Nighttime light data and Economic growth

The application earthly satellite images at night (night lights data) was initially invented to study human activities and natural events. However, the idea was extended to economics, arguing that night lights can be good measures of economic growth for any active economy. Additionally, night lights can be used to measure poverty, and social inequalities. Also, night lights data helps in answering questions that previously had no answers especially in places with insufficient data. To deeply understand the role of night lights data on economic growth, we briefly look at the composition of the satellite image pixels and their economic relevancy. Ideally, each pixel of a satellite image represents an area of less than one square kilometer on earth. It is comprising of a digital number measuring the brightness at night. Intuitively, the higher the number the brighter the spot. When we integrate over all pixels for a given country (South Africa), it becomes an indicator that measures the activities of that country at night. In this study, we integrate these numbers over all pixels for industrial sectors with insurance businesses for South Africa. This gives us a barometer of economic development and fluctuations in South Africa. Ideally, night lights help in measuring the dynamics and developmental changes of one or more economies over time, yet GDP cannot capture this phenomenon in depth.

## 2. Methods

### 2.1. Data and variables

We used time series South African GDP data collected from the World Bank for the period running from 2000 to 2018. Our choice for South Africa is entirely driven by the motive to assess the economic growth patterns and rate for emerging economies. Additionally, we used a fine-grained geo-coded industrial full sample micro-data set for South Africa and match it with radiance and saturated light emissions. We collected nighttime lights data from the National Geophysical Data Centre (NGDC) in partnership with the National Oceanic and Atmospheric Administration (NOAA). The data is used to test our set of hypotheses. Overall, our analysis consulted the following variables:

| Variable | Description |
| --- | --- |
| Insurance Presence | A measure of points where insurance entities are located (authors' definition) |
| Insurance Count | The total number of insurance firms in any geographical area, (authors' definition) |
| Mean_light | The average of the nighttime light generated data, Mellander et al (2016) |
| Sum_light | The sum(total) of the nighttime light generated data, Mellander et al (2016) |
| GDP/Capita | The total output produced in an economy measured per head, OECD(2023) |
| GDP | Gross domestic output produced within the South African borders, OECD (2023) |
| Insurance premiums | The amount paid by policyholders to insurers for insurance coverage every month, Olayungbo & Akinlo,(2016) |

| Insurance density | It is an indicator of the development of insurance within a country (South Africa) and the ratio of total insurance premiums to the total population, Olayungbo & Akinlo,(2016) |

Table 1: Variable description.

## 2.2. Matching the Nighttime lights data (NTL) and Economic Data

To compare the light data with the Statistics of South African demographic data, we used a generated grid based on the coordinates of the provided data on the concentration of the insurance companies and matched geographical locations and respective economic outputs. The light data were then summarized in the grid to allow for analysis. Using ESRI's ArcGIS software, we created a grid of squares based on the lower left coordinates provided by the Statistics department of South Africa. Separate grids were generated for the insurance industry and population data, resulting in two grids overall (the coordinates provided were based on the available data). This was to refine our analysis and to make them precise. In addition, we regressed real GDP/Capita and nighttime lights (see results).

## 2.3. Correlation analysis

We now turn to the empirical analysis, connecting the light-emission data, GDP data, and the selected insurance indicators. We begin with a fundamental correlation analysis to identify any bivariate relations. Separate analyses were done on (1) the correlation of activity light intensity and insurance premium density and (2) GDP/capita and insurance premium density. We employed Radiance light and Saturated Light, GDP/capita, GDP, and insurance indicators in the correlations for robust comparisons. We also computed correlations for specific industry and occupational structures, with all correlation coefficients significantly weaker than for total people or industry values. However, the correlation between people and light is generally more robust when we use radiance light emissions instead of saturated light. Also, the correlation between GDP, GDP/capita, and insurance indicators are weaker than for nighttime data.

### 2.4. Weighted Regression(WR) Analysis

Weighted regression is a technique that allows us to examine possible spatial non-stationarity by using weighted sub-samples of the data. This implies that we can produce locally linear regression estimates for every point in space. In other words, we generate $\beta$, which is a local estimate for every observation. This implies that the $\beta$ will be there to capture any effect on the systematic estimates. The effect can be geographical such as latitude changes. Such changes can, in some way, affect economic activity. These disparities will then be captured by the Beta coefficients in the selected regions of South Africa and visible in the maps generated from the analysis. The methodology makes it possible to compare the unstandardized $\beta$ coefficients from the OLS estimation of the involved parameters. In simpler terms, the idea is that the WR estimation produces information about parameter variation over space. The difference between WR and spatial autocorrelation techniques is that the latter identifies spatial dependence through the residual, while the former addresses spatial non-stationarity directly through the estimated parameters. In a WR, we assume the regression model to be:

$$y_i = \beta_0(i) + \beta_1(i)x_{i1} + \beta_2(i)x_{i2} + \cdots + \beta_t(i)x_{it} \ldots + \epsilon_{it,,,,,,,,,(1)}$$

The estimator of our parameters is given by;

$$\hat{\beta}(i) = (X^T V(i) X)^{-1} X^T V(i) Y ,,,,,,,,,,,,,, (2)$$

Where $V(i)$ is a weight specific matrix to location $i$, observations near to $i$ are given greater weight than distant observations. However, in our empirical analysis below, all estimations will be in single regressions. This is done to comply with the strong multicollinearity between the independent variables. $\hat{\beta}(i)$ is the vector of parameters. These parameters are sensitive and form the basis of our WR models. $X$ is the vector specifying our covariates incorporated in this study, and they explain South African economic development. This study incorporates variables like insurance premiums and insurance density. $Y$ is the dependent economic development variable, which we seek to explore its movements against insurance sectorial developments. $\epsilon_i$ is the error term. Ideally, the error term captures the residuals from the WR model, which are other factors contributing to economic development but excluded from our current model.

We aim to examine if we can find any strong interdependencies between economic development and the insurance sector where the economic growth measurements are compared. In case of failure of the radiance emissions generated somewhat more robust correlation coefficients, the radiance light emission is used instead of the saturated light data as the dependent variable or our economic measure in all regressions below. Also, GDP and GDP per capita shall be used as control variables. Our regressions are run in the log-log functional form to cater to contemporaneous correlations.

## 2.5. Inter Province-Annual variation in GDP/Capita relative to Insurance industry contribution

To present a simple informative trend analysis of the insurance sector growth in South Africa, we provide a proportionality growth analysis below.

| Province | GDP/Capita(%) | | | | | | | | | | | | | |
|---|---|---|---|---|---|---|---|---|---|---|---|---|---|---|
| | 2008 | 2009 | 2010 | 2011 | 2012 | 2013 | 2014 | 2015 | 2016 | 2017 | 2018 | 2019 | 2020 | 2021 |
| Gauteng | 33.92% | 33.97% | 34.10% | 34.24% | 34.40% | 34.50% | 34.68% | 34.66% | 34.94% | 34.82% | 34.94% | 35.08% | 35.01% | 34.96% |
| KwaZulu-Natal | 15.68% | 15.71% | 15.79% | 15.85% | 15.91% | 15.92% | 16.01% | 15.97% | 15.99% | 16.05% | 16.04% | 16.03% | 16.04% | 16.04% |
| Western Cape | 13.61% | 13.63% | 13.57% | 13.63% | 13.71% | 13.72% | 13.77% | 13.81% | 13.88% | 13.86% | 13.86% | 13.89% | 13.88% | 13.87% |
| Eastern Cape | 7.71% | 7.75% | 7.70% | 7.73% | 7.72% | 7.64% | 7.59% | 7.57% | 7.59% | 7.53% | 7.51% | 7.50% | 7.51% | 7.51% |
| Mpumalanga | 7.41% | 7.42% | 7.38% | 7.31% | 7.30% | 7.27% | 7.34% | 7.24% | 7.22% | 7.25% | 7.24% | 7.20% | 7.22% | 7.23% |
| Limpopo | 7.39% | 7.39% | 7.36% | 7.30% | 7.21% | 7.22% | 7.17% | 7.21% | 7.15% | 7.20% | 7.18% | 7.16% | 7.17% | 7.18% |
| North West | 6.64% | 6.55% | 6.56% | 6.49% | 6.24% | 6.26% | 5.92% | 6.11% | 5.87% | 5.90% | 5.89% | 5.85% | 5.87% | 5.88% |
| Free State | 5.43% | 5.38% | 5.36% | 5.29% | 5.33% | 5.30% | 5.31% | 5.23% | 5.19% | 5.20% | 5.14% | 5.11% | 5.13% | 5.15% |
| Northern Cape | 2.22% | 2.20% | 2.19% | 2.16% | 2.18% | 2.18% | 2.20% | 2.20% | 2.17% | 2.20% | 2.19% | 2.17% | 2.18% | 2.19% |

| Province | Insurance Contribution(%) | | | | | | | | | | | | | |
|---|---|---|---|---|---|---|---|---|---|---|---|---|---|---|
| | 2008 | 2009 | 2010 | 2011 | 2012 | 2013 | 2014 | 2015 | 2016 | 2017 | 2018 | 2019 | 2020 | 2021 |
| Gauteng | 1.25% | 1.36% | 1.41% | 1.45% | 1.50% | 1.54% | 1.59% | 1.65% | 1.72% | 1.78% | 1.84% | 1.91% | 1.84% | 1.69% |
| KwaZulu-Natal | 0.35% | 1.00% | 0.95% | 0.49% | 0.76% | 0.78% | 1.08% | 0.78% | 0.99% | 0.63% | 0.71% | 0.44% | 0.60% | 0.87% |
| Western Cape | 0.50% | 0.76% | 0.40% | 0.46% | 0.36% | 1.10% | 1.03% | 0.87% | 0.61% | 0.39% | 1.02% | 0.73% | 0.71% | 0.72% |
| Eastern Cape | 0.90% | 1.09% | 1.02% | 0.82% | 0.87% | 0.96% | 0.48% | 0.88% | 0.89% | 0.29% | 1.02% | 0.90% | 0.74% | 0.64% |
| Mpumalanga | 0.26% | 0.57% | 0.86% | 0.07% | 1.04% | 0.89% | 0.70% | 0.17% | 1.33% | 1.04% | 0.53% | 0.94% | 0.84% | 0.81% |
| Limpopo | 0.96% | 0.56% | 1.56% | 0.34% | 0.93% | 0.81% | 0.40% | 0.58% | 0.29% | 1.17% | 0.99% | 0.98% | 1.05% | 0.61% |
| North West | 0.85% | 0.73% | 1.37% | 1.07% | 1.10% | 1.20% | 0.55% | 0.57% | 0.33% | 0.39% | 0.57% | 1.00% | 0.65% | 0.46% |
| Free State | 0.56% | 1.25% | 0.80% | 0.88% | 0.35% | 0.60% | 0.67% | 0.93% | 0.89% | 0.81% | 0.37% | 1.19% | 0.79% | 0.83% |
| Northern Cape | 1.68% | 0.60% | 1.05% | 1.16% | 1.00% | 0.39% | 0.70% | 1.55% | 0.52% | 0.90% | 0.93% | 0.86% | 0.90% | 0.92% |

Table 2: Proportion of variation in the insurance industry at the provincial level in South Africa (2008-2021).

The table above shows the variations associated with insurance sector growth and corresponding real GDP/Capita at a provincial level in South Africa for the period spanning from 2008 to 2021. Nine provinces are considered as indicated in the table above, and provinces with high insurance activity (growth percentage) have corresponding high real GDP/Capita over the selected period, which in this case is Gauteng Province. The table provides us with necessary information on the role of the insurance sector on economic growth before running our in-depth regressions.

## 3. Results and Findings

We start by summarizing statistics on the main variables included in this study. We reported each variable's mean, standard deviation, and minimum and maximum values.

### 3.1. Descriptive statistics

| Variable | Min | Max | Mean | Std. Deviation |
|---|---|---|---|---|
| Insurance Presence | - | 967.00 | 640.03 | 73.86 |
| Radiance light intensity** | - | 85.00 | 12.50 | 17.31 |
| Saturated light Intensity | - | 105.00 | 18.95 | 21.34 |
| GDP/Capita(%)** | - | 2.75 | 1.33 | 4.37 |
| GDP(%) | - | 4.10 | 2.13 | 5.91 |
| Insurance premium density(premium/Insured Population)** | - | 6,752.00 | 55,293.00 | 12,771.00 |
| Insured Population Density(insured/total polution) | - | 97,663.00 | 527.34 | 1,994.00 |
| N | 155,168.00 | | | |

**Comments**

Table 3 shows the summary statistics of all the variables used in this study. N is the number of sample size. Relative to radiance light, saturated light has high mean and standard deviation values of 18.95 and 21.34, respectively, whereas the corresponding values for radiance light are 12.5 and 17.31. Radiance light and GDP/Capita shall be compared since GDP has a greater mean and standard deviation of 2.13 and 5.91, respectively, than GDP/Capita. A significant mean implies the skewness of the data either positively or negatively, and it deviates from normality and is true otherwise. Thus the data variable with a small mean is considered with a small corresponding standard deviation. GDP/Capita and Radiance light are used and compared accordingly.

### 3.2. Correlation analysis

Below is a matrix for the correlation coefficients across all the variables used in this paper.

**Correlation Coefficients (Pearson correlation)**

| Variable | Insurance presence | Insurance density | Radiance light | Saturated light | GDP | GDP/Capita |
|---|---|---|---|---|---|---|
| Insurance presence | 1 | | | | | |
| Insurance density | 0.34 | 1 | | | | |
| Radiance light | 0.936 | 0.8837 | 1 | | | |
| Saturated light | 0.7336 | 0.605 | 0.2054 | 1 | | |
| GDP | 0.7561 | 0.5114 | 0.656 | 0.5530 | 1 | |
| GDP/Capita | 0.8314 | 0.772 | 0.944 | 0.6209 | 0.9257 | 1 |

Table 4: Correlation matrix for the variables used in this study. **Comments**

We examined if there exist any relationships among all our variables of interest. We employ both Radiance light as well as Saturated light in the correlations to be able to examine possible differences. From table 2 above, the pair correlations across all the variables are strongly positive.

When measuring insurance presence and density against either GDP/Capita and Radiance light (our core variables of interest), we observe that insurance density and presence are more highly correlated to radiance light than GDP/Capita, indicating its proxy efficiency in measuring economic activity. Although GDP/Capita and radiance light tend to move in the same direction, we further investigated it using the scatter plot presented below.

**Nighttime lights data versus GDP per Capita**

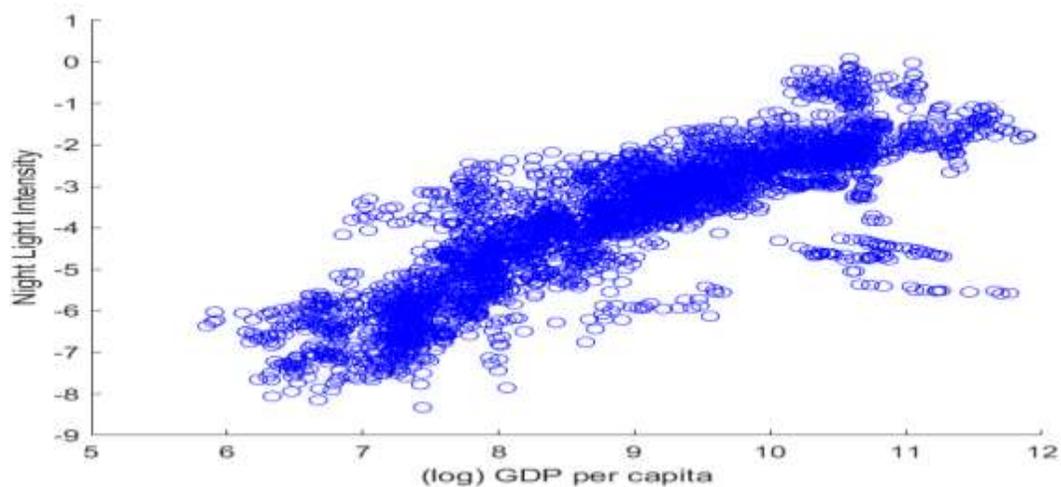

Figure 4: Scatter plot for Radiance light intensity and GDP/Capita based on log transformations.

A simple informative scatter plot above depicts some contemporaneous correlations between GDP/Capita and Nighttime light intensity in South Africa. Although there is a positive correlation between the two, nighttime light intensity (radiance light) will soon prove to be a proxy when measuring economic growth. Meanwhile, the nighttime lights are related to the true latent GDP/Capita through an unknown production function $m(\cdot)$ and an additive error term. Mathematically, this can be presented as below:

$$Z_{i,t} = m(y_{i,t}^*) + \varepsilon_{i,t}^z(I_i) \quad (3)$$

The error term distribution above varies with the geographic locations; see Hu and Yao (2019). The data inform the specification of the production function $m(\cdot)$. Logarithmic transformation is applied to the GDP data to enhance a good fit. We regress GDP/Capita on nighttime lights for South Africa (its provinces) with year dummies. We conjectured that Nighttime lights predict real GDP per capita.

### 3.3. Weighted Regression analysis

We ran weighted regressions, incorporating all our variables, with radiance light as the dependent variable. Ordinary least squares (OLS) add less value to geographically weighted data and are more or less similar to the results obtained from the above correlations. Also, weighted regressions help control multicollinearity problems among our variables. For example, the variable insurance presence is omitted and dropped out of the model due to collinearity. In this study, we assume our WR model to be

$$y_i = \beta_0(i) + \beta_1(i)x_{1i} + \cdots + \beta_k(i)x_{ki} + \varepsilon_i \ldots\ldots\ldots\ldots (6)$$

or equivalently $Y = X\beta + \varepsilon$

The estimator is given by:

$$\acute{\beta}(\iota) = (X^T W(i) X)^{-1} X^T W(i) Y \ldots\ldots\ldots\ldots (7)$$

Where $W(i)$ is a weight-specific matrix to location $i$, so that observations near to $i$ are given greater weight than the remote ones, $X$ is the matrix of our regressors/independent and predetermined variables. $Y$ is the dependent variable matrix. The results are in the table below.

| Variable | No of Observations | Min | Lowerquartile | Mean | Global_OLS | Upper quartile | Max | Neighbors |
|---|---|---|---|---|---|---|---|---|
| Insured Population Density(insured/total polution) | 155,168 | -0.366 | 0.055 | 0.172 | 0.544 | 0.261 | 0.963 | 116.000 |
| Radiance light intensity** | 155,168 | -0.075 | 0.090 | 0.170 | 0.424 | 0.226 | 0.908 | 407.000 |
| GDP(%) | 155,168 | -0.022 | 0.165 | 0.231 | 0.400 | 0.294 | 0.599 | 639.000 |
| Insurance Presence | 155,168 | -2.685 | -0.010 | 0.088 | 0.944 | 0.115 | 3.649 | 30.000 |
| GDP/Capita(%)** | 155,168 | -0.612 | 0.032 | 0.171 | 0.692 | 0.257 | 1.814 | 62.000 |
| Insurance Presence | 155,168 | -0.695 | -0.002 | 0.045 | 0.420 | 0.060 | 1.684 | 38.000 |
| Saturated light Intensity | 155,168 | -0.241 | 0.014 | 0.091 | 0.417 | 0.136 | 1.168 | 69.000 |
| Insurance premium density(premium/Insured Population)** | 155,168 | -0.270 | 0.001 | 0.031 | 0.176 | 0.045 | 0.495 | 87.000 |

| Variable | R^2 | OLS_AIC | GWR_AIC | Residual Squares |
|---|---|---|---|---|
| Insured Population Density(insured/total polution) | 0.582 | 568, 130.7 | 239680.0 | 35784.3 |
| Radiance light intensity** | 0.275 | 672, 166 | 459727.3 | 122551.2 |
| GDP(%) | 0.490 | 605692.4 | 453967.4 | 120,111.8 |
| Insurance Presence | 0.240 | 417341.5 | 98928.7 | 10370.5 |
| GDP/Capita(%)** | 0.573 | 350702.4 | 115,807 | 15,311.45 |
| Insurance Presence | 0.226 | 419521.0 | 113,875.5 | 13074.6 |
| Saturated light Intensity | 0.461 | 377607.3 | 141256.6 | 19,401.43 |
| Insurance premium density(premium/Insured Population)** | 0.208 | 422, 154.3 | 185909.5 | 29480.3 |

Table 5: All regressions are in the log-log format.

**Comments**

From both the OLS and Geographically weighted regression (GWR) performed, we obtain a low AIC value from GWR than OLS, indicating its superiority when investigating variations across geographical space. An adaptive approach is used to compute the GWR model parameter estimates based on the weighted neighborhood optimization trickery. Insurance density has the highest spatial impact on the regression parameter estimates others (with radiance light as the dependent variable). When using GDP/Capita, we observe very minimal sensitivity indicating the superiority of night lights as a measure of economic activity. This is consistent with the results provided earlier on correlation analysis.

## 4. Analysis of time series Nighttime light and statistical GDP/Capita Data

The urban light affected the unlit regions outside the South African Insurance industry. Therefore, $DN$ values are more extensive than 0 for the shady area in the time series DMSP-OLS imageries. Following the work by Sutton et al. (2007), we selected a $DN$ value threshold of 40 for nighttime light imageries to delimit urban extents. We used 40 due to the wide geographical separation of the South African cities. We divided South Africa into six economic regions and determined the optimal thresholds for each economic region, which are 36 – 68. We believe these thresholds are sufficient for South Africa, a good emerging economic country. However, GDP and GDP/head come from not only cities but also the countryside. Those significant thresholds will ignore light information from small cities and the countryside, which also contribute to GDP when the threshold is too large. The blooming effects of the imageries will be hard to minimize, and some regions without light are also counted when the selected threshold is too small. A $DN$ value threshold of 10 was selected to minimize the effects of blooming. We adopted the idea Zhao et al. (2012) discovered in their China-based study. The light blooming effects controls are as follows:

$DN_{adj} = \begin{pmatrix} 0, & DN < 0 \\ DN, & DN \geq 0 \end{pmatrix}$ ,,,,,,,,,,,, (10), where $DN_{adj}$ is the $DN$ values in nighttime imageries after adjusting, and *DN* indicates the DN values in the original imageries. We combined GDP and Nighttime light data to obtain GDP per unit light intensity. We assumed that the insurance industry in South Africa had been developed to produce more intense or comprehensive lights when the intensity of Nighttime light increases rapidly relative to GDP in one year. We analyzed the time series province-level GDP per unit light intensity to investigate the variation trend of the insurance industry structure in different provinces. RGDP is given as follows;

$$R_{GDP} = \frac{GDP_i}{DN_i}$$

where $R_{GDP}$ indicates the GDP per unit of light intensity, $GDP_i$ indicates provincial GDP at the time $i$, and $DN_i$ indicates the sum of DN of the $i^{th}$ province. This study, therefore, investigated the correlation between variations in the proportion of the insurance industry and $R_{GDP}$ based on a comparative analysis of time series $R_{GDP}$ and the annual proportion of the insurance industry (measured by insurance premium density) in each province defined accordingly.

**Model validation**

| Model | Likelihood ratio test | BIC | AIC |
|---|---|---|---|
| Model 1(GDP) | 15.45% | 45.8% | 42.14% |
| Model 2(GDP/Capita) | 23.75% | 37.87% | 31.09% |
| Model 3(Radiance light) | 2.13% | 4.14% | 6.71% |
| Model 4(Saturated light) | 4.31% | 6.45% | 5,01% |

Table 6presents is a comparison and assessment of the models employed in this study. Ideally, we wanted to choose the optimal and best model which closely and accurately measures economic growth and activities. From the table above, Model 3 outperformed the rest, as indicated by small values of the BIC, AIC, and likelihood ratios. Interestingly, model 1 and model 2, which used GDP performed worst. This concludes that nighttime light data is a good measure of economic activities in South Africa.

## 5. Discussions and Conclusions

Economists are still working to extract the proper signal from looking at satellite photographs. We must better understand why the changes in lights in regions and countries over time do not perfectly predict the changes in their total economic activity. We may have to wait for new sensors, such as the Visible Infrared Imaging Radiometer Suite (VIIRS), to get more precise light readings from places on earth. Nevertheless, the satellite photographs will give us consistent monitoring of where economies grow or decline. Such detailed knowledge will guide researchers and policymakers on what helps development and what policies we need to promote it. This study looks at insurance growth impacts on economic growth and aims to determine whether nighttime light data is a proxy measure of economic growth than GDP and GDP/Capita. Radiance light instead of saturated light is used to measure nighttime light activities.

From the results obtained, nighttime light data is a good proxy measure of economic growth than either GDP or GDP/Capita. Nighttime light data is a better proxy measure of economic growth than GDP and GDP/capita. As such, policymakers and government, not only of South Africa, should pay attention to this aspect. Also, the insurance sector is vital to economic growth and development, and particular focus and attention should be paid, such as funds disbursement and sectorial subsidization. Overall, the non-banking sector (insurance industry) drives economic growth, and nighttime light data is a good proxy of economic growth in economies with poor statistical protocols.

While our study addressed a fundamental economic development question from an insurance sector and night-time lights data perspectives, we note that the use of night-time lights data is costly and time intensive. For less biased results and reusable conclusions, it requires economic and quantitative commitment and this is a challenge in developing economies. However, given adequate financial and tech based equipment, the results from the methodology and approach are phenomenal. Despite these limitations, we also note that the findings will guide researchers and policymakers on what drives economic development and what policies to put in place, as well as efficient resource allocation. It would be interesting to extend the current study to other non-banking sectors such as micro-finances, mutual and hedge funds.